\documentclass[preprint,amsmath,amssymb,aps,showkeys,showpacs]{revtex4}
\usepackage[english]{babel}
\usepackage{graphicx}
\usepackage{graphics}
\usepackage{amsmath}
\usepackage{dcolumn}
\usepackage{amssymb}
\usepackage{bm}
\usepackage[latin1]{inputenc}

\begin{document}
\title{Relativistic Landau-He-McKellar-Wilkens quantization and relativistic bound states solutions for a Coulomb-like potential induced by the Lorentz symmetry breaking effects}
\author{K. Bakke}
\email{kbakke@fisica.ufpb.br}
\affiliation{Departamento de F\'isica, Universidade Federal da Para\'iba, Caixa Postal 5008, 58051-970, Jo\~ao Pessoa, PB, Brazil.}

\author{H. Belich} 
\affiliation{Departamento de F\'{\i}sica e Qu\'{\i}mica, Universidade Federal do Esp\'{\i}rito Santo, Av. Fernando Ferrari, 514, Goiabeiras, 29060-900, Vit\'{o}ria, ES, Brazil.}

\begin{abstract}
In this work, we discuss the relativistic Landau-He-McKellar-Wilkens quantization and relativistic bound states solutions for a Dirac neutral particle under the influence of a Coulomb-like potential induced by the Lorentz symmetry breaking effects. We present new possible scenarios of studying Lorentz symmetry breaking effects by fixing the space-like vector field background in special configurations. It is worth mentioning that we use the criterion for studying the violation of Lorentz symmetry is that the gauge symmetry should be preserved. 
\end{abstract}

\maketitle

\section{Introduction}

One of the most known quantum system in the literature arises from the interaction between a charged particle and a uniform magnetic field yielding discrete energy levels. These discrete energy levels corresponds to the quantization of a charged particle in cyclotron orbits which is called the Landau quantization \cite{landau1,landau}. In recent decades, the Landau quantization has been investigated in studies of Bose-Einstein condensate \cite{l1}, quantum Hall effect \cite{l2}, linear topological defects \cite{l3,l4} and for neutral particles \cite{er,lin,b}. The Landau quantization was also extended to studies of relativistic quantum mechanics by Rabi \cite{l5}, Berestetskii {\it et al.} \cite{l6}, Jackiw \cite{l7} and Balatsky {\it et al.} \cite{l8}. The interest in the relativistic Landau quantization has attracted discussions about spin nematic state \cite{l9}, the finite-temperature problem \cite{l10} and quantum Hall effect \cite{l11}. In recent years, the relativistic Landau quantization has been extended to neutral particles \cite{bf5}, and it has been discussed, in a noninertial reference frame \cite{b4} and in the Lorentz symmetry violation background \cite{bbs}. In Dirac-like systems, the Landau quantization has been investigated under the influence of topological defects in \cite{l12}.

The aim of this work is to investigate the relativistic Landau-He-McKellar-Wilkens quantization and bound states solutions for a Dirac neutral particle under the influence of a Coulomb-like potential induced by the Lorentz symmetry breaking effects. A fundamental guide in the study of any theory is the symmetry because it deals with conservation laws. For instance, the Lorentz and CPT symmetries have a supreme importance in Quantum Field Theory since the the Standard Model for Particle Physics is based on both of them. Another point of view involving symmetries in physical systems corresponds to the symmetry breaking. In nonrelativistic quantum systems, well-known quantum effects related to symmetry breaking are phase transitions in ferromagnetic systems, where the rotation symmetry is broken when the system is under the influence of a magnetic field. The study of symmetry breaking can also be extended to relativistic systems by considering a background given by a $4$-vector field that breaks the symmetry $\mathcal{SO}\left( 1,3\right)$ instead of the symmetry $\mathcal{SO}\left( 3\right)$. This line of research is known in the literature as the spontaneous violation of the Lorentz symmetry \cite{extra3,extra1,extra2}. 

Kostelecky and Samuel \cite{extra3} were the first authors to suggest the study of this new possibility of spontaneous violation by observing the spontaneous violation of symmetry by a scalar field could be extended in the string field theory. As a result of this spontaneous violation we have a ``four-vector'' background that, in fact, behaves as four scalar components by a Lorentz transformation. These components are Lorentz invariants, and do not transform by active Lorentz transformations. However, this behaviour is not covariant, therefore the Lorentz symmetry is violated in this sense \cite{eu}. Furthermore, in the electroweak theory, a scalar field acquires a nonzero vacuum expectation value which yields mass to gauge bosons (the Higgs Mechanism). Recently, theories involving spontaneous violation of symmetry are encompassed in the framework of the Standard Model Extension \cite{col} as a possible extension of the minimal Standard Model of the fundamental interactions. As example, the violation of the Lorentz symmetry is implemented in the fermion section of the Standard Model Extension by two CPT-odd terms: $v_{\mu}\overline{\psi}\gamma^{\mu }\psi $ and $b_{\mu}\overline{\psi}\gamma _{5}\gamma^{\mu}\psi $, where $v_{\mu}$ and $b_{\mu}$ correspond to the Lorentz-violating backgrounds. An interesting discussion made in Ref. \cite{shap} deals with the $b_{\mu}$ field as a constant torsion field. From this perspective, it is estimated the fermion-torsion interaction is under the limit of $\left|\vec{b}\right|<10^{-30}\,\mathrm{Gev}$. Other aspects of the Standard Model Extension that have been examined in recent years are supersymmetry \cite{Susy}, Casimir effect \cite{Casimir}, radiative corrections \cite{Radiative}, vacuum Cherenkov radiation emission \cite{Cherenkov2}, and anisotropies of the Cosmic Microwave Background Radiation \cite{CMBR}.

In recent years, neutrino oscillations have established a real necessity of discussing physics beyond the Standard Model. Although the Standard Model is believed to be a low-energy effective theory from a unified description of gravity and quantum physics at the Planck scale, a spontaneous violation of the Lorentz symmetry can occurs beyond the electroweak scale. Thereby, a theory of quantum gravity necessarily must include such violation effects, then, one can analyse neutral fermions in a general coordinate system in order to investigate the influence of the Lorentz violation background on a moving neutrino \cite{atual}. Therefore, we propose the study of Lorentz symmetry breaking effects through a classical background such as $v_{\mu}\overline{\psi}\gamma^{\mu }\psi$ and $b_{\mu}\overline{\psi}\gamma _{5}\gamma^{\mu}\psi$ as a mean field theory, hence it does not have the commitment to explain the microscopic mechanism. This behaviour is quite unusual because it is due to a vacuum contribution, i. e., a neutral particle feels such coupling in presence of this environment as a groundstate of a fundamental theory. This line of research proposes, as evidence, different scenarios in which the spontaneously Violation of Lorentz symmetry \cite{Nonmini} could occur.

Other studies of symmetry breaking were done by modifying the Dirac theory \cite{Hamilton}. Nowadays, there exists a great deal of works looking at Lorentz violation, and numerous experimental bounds \cite{extra2}. In the nonrelativistic limit of the modified Dirac theory, the spectrum of energy of the hydrogen atom was discussed in \cite{Nonmini}. By introducing nonminimal couplings into the Dirac equation in order to describe new possible scenarios of the Lorentz symmetry violation background, geometric quantum phases \cite{anan2,berry,zei2} have been studied based on Lorentz symmetry breaking effects \cite{belich,belich1,belich2,belich3,bbs2,bbs3,bb} in recent years.

In this paper, we study the relativistic Landau-He-McKellar-Wilkens quantization and bound states solutions for a Dirac neutral particle under the influence of a Coulomb-like potential induced by the Lorentz symmetry breaking effects. We present two possible scenarios of studying Lorentz symmetry breaking effects: the first relativistic scenario is based on the assumption of a fixed space-like vector field background parallel to the $z$ axis of the spacetime and a radial magnetic field produced by a magnetic charge density; the second relativistic scenario is based on the assumption of a fixed space-like vector field background parallel to the radial direction and a uniform magnetic field on the $z$ direction. It is worth mentioning that we use the criterion for studying the violation of Lorentz symmetry is that the gauge symmetry should be preserved.

This paper is organized as follows: in section II, we present the Lorentz symmetry violation background and discuss the way of achieving the relativistic Landau-He-McKellar-Wilkens quantization induced by Lorentz symmetry breaking effects; in section III, we present another background of the Lorentz symmetry violation and obtain bound state solutions for the Dirac equation from the influence of a Coulomb-like potential induced by the Lorentz symmetry breaking effects; in section IV, we present our conclusions.

\section{relativistic Landau-He-McKellar-Wilkens quantization induced by Lorentz symmetry breaking effects}

In this section, we discuss the relativistic Landau-He-McKellar-Wilkens quantization induced by a Lorentz symmetry violation background. The relativistic Landau-He-McKellar-Wilkens quantization was proposed in \cite{bf5} by employing the duality transformation on the relativistic Landau-Aharonov-Casher system.
%equations of motion which describe the relativistic Landau-Aharonov-Casher quantization. 
Based on this study, the relativistic Landau-He-McKellar-Wilkens can be achieved by imposing that the field configuration that interacts with a neutral particle satisfies the electrostatic conditions, produces no torque on the electric dipole moment, and yields the presence of an effective uniform magnetic field given by $\vec{B}_{\mathrm{eff}}=\vec{\nabla}\times\left(\vec{n}\times\vec{B}\right)$ (with $\vec{n}$ being the direction of the permanent electric dipole moment of the neutral particle). Recently, the relativistic Landau-He-McKellar-Wilkens has been studied in a noninertial reference frame \cite{b4}. In the following, we present the Lorentz symmetry violation background which allows us to obtain the analogue of the relativistic Landau-He-McKellar-Wilkens quantization. 

We start by introducing the Lorentz symmetry violation background. Recently, we have proposed the study of the He-McKellar-Wikens effect \cite{hmw} and the scalar Aharonov-Bohm effect \cite{zei2,anan2} based on Lorentz symmetry breaking effects by modifying the nonminimal coupling proposed in \cite{belich1,belich} and writing it in the form:
\begin{eqnarray}
i\gamma^{\mu}\partial_{\mu}\rightarrow i\gamma^{\mu}\partial_{\mu}-g\,b^{\mu}\,F_{\mu\nu}\left(x\right)\,\gamma^{\nu},
\label{1}
\end{eqnarray}
where $g$ is a constant, $b^{\mu}$ corresponds to a fixed $4$-vector that acts as a vector field breaking the Lorentz symmetry violation, the tensor $F_{\mu\nu}\left(x\right)$ corresponds to the usual electromagnetic tensor ($F_{0i}=-F_{i0}=-E_{i}$, and $F_{ij}=-F_{ji}=\epsilon_{ijk}B^{k}$), and the $\gamma^{\mu}$ matrices are defined in the Minkowski spacetime in the form \cite{greiner}:
\begin{eqnarray}
\gamma^{0}=\hat{\beta}=\left(
\begin{array}{cc}
1 & 0 \\
0 & -1 \\
\end{array}\right);\,\,\,\,\,\,
\gamma^{i}=\hat{\beta}\,\hat{\alpha}^{i}=\left(
\begin{array}{cc}
 0 & \sigma^{i} \\
-\sigma^{i} & 0 \\
\end{array}\right);\,\,\,\,\,\,\Sigma^{i}=\left(
\begin{array}{cc}
\sigma^{i} & 0 \\
0 & \sigma^{i} \\	
\end{array}\right),
\label{2}
\end{eqnarray}
with $\vec{\Sigma}$ being the spin vector. The matrices $\sigma^{i}$ correspond to the Pauli matrices, and satisfy the relation $\left(\sigma^{i}\,\sigma^{j}+\sigma^{j}\,\sigma^{i}\right)=2\eta^{ij}$. Our interest is to work with curvilinear coordinates, thus, we need to apply a coordinate transformation $\frac{\partial}{\partial x^{\mu}}=\frac{\partial \bar{x}^{\nu}}{\partial x^{\mu}}\,\frac{\partial}{\partial\bar{x}^{\nu}}$, and a unitary transformation on the wave function $\psi\left(x\right)=U\,\psi'\left(\bar{x}\right)$ \cite{schu,bf5,bbs3}. In this way, the Dirac equation can be written in any orthogonal system in the presence of Lorentz symmetry breaking effects described in (\ref{1}) as \cite{bbs3} 
\begin{eqnarray}
i\,\gamma^{\mu}\,D_{\mu}\,\psi+\frac{i}{2}\,\sum_{k=1}^{3}\,\gamma^{k}\,\left[D_{k}\,\ln\left(\frac{h_{1}\,h_{2}\,h_{3}}{h_{k}}\right)\right]\psi-g\,b^{\mu}\,F_{\mu\nu}\left(x\right)\,\gamma^{\nu}\psi=m\psi,
\label{4}
\end{eqnarray}
where $D_{\mu}=\frac{1}{h_{\mu}}\,\partial_{\mu}$ is the derivative of the corresponding coordinate system, and the parameter $h_{k}$ correspond to the scale factors of this coordinate system \cite{schu}. For instance, the line element of the Minkowski spacetime is writing in cylindrical coordinates in the form: $ds^{2}=-dt^{2}+d\rho^{2}+\rho^{2}d\varphi^{2}+dz^{2}$; then, the corresponding scale factors are $h_{0}=1$, $h_{1}=1$, $h_{2}=\rho$ and $h_{3}=1$. Moreover, the second term in (\ref{4}) gives rise to a term called the spinorial connection $\Gamma_{\mu}\left(x\right)$ \cite{schu,b4,bbs,bbs2,bbs3,weinberg}. In this way, the Dirac equation (in cylindrical coordinates) in the Lorentz symmetry violation background described by the nonminimal coupling (\ref{1}) is given by \cite{bbs3}
\begin{eqnarray}
i\frac{\partial\psi}{\partial t}=m\hat{\beta}\psi+\vec{\alpha}\cdot\left[\vec{\pi}-g\,b^{0}\vec{E}-g\,\left(\vec{b}\times\vec{B}\right)\right]\psi+g\,\vec{b}\cdot\vec{E}\psi,
\label{5}
\end{eqnarray}
where $\pi_{k}=-i\partial_{k}-i\Gamma_{k}$ and $\Gamma_{\varphi}=-\frac{i}{2}\,\Sigma^{3}$. Note that we can define an effective potential vector $\vec{A}_{\mathrm{eff}}=b^{0}\vec{E}+\left(\vec{b}\times\vec{B}\right)$ in such a way that there exists an effective magnetic field given by
\begin{eqnarray}
\vec{B}_{\mathrm{eff}}=\vec{\nabla}\times\left[b^{0}\,\vec{E}+\left(\vec{b}\times\vec{B}\right)\right].
\label{5a}
\end{eqnarray}

Now, we are able to discuss the relativistic Landau-He-McKellar-Wilkens quantization in a Lorentz symmetry violation background. First of all, we consider a space-like fixed vector and a field configuration given by
\begin{eqnarray}
\vec{b}=b^{3}\,\hat{z};\,\,\,\,\,\,\vec{B}=\frac{\lambda_{m}\rho}{2}\,\hat{\rho},
\label{6}
\end{eqnarray}
where $\lambda_{m}$ corresponds to a magnetic charge density \cite{lin,bf5}, and the vectors $\hat{z}$ and $\hat{\rho}$ correspond to the unit vectors in the $z$-direction and the radial direction, respectively. Note that, based on the Lorentz symmetry violation background (\ref{6}), we can consider the electric dipole moment of the neutral particle as $\vec{d}=g\,\vec{b}$, then, we have an effective potential vector $\vec{A}_{\mathrm{eff}}'=\left(\vec{b}\times\vec{B}\right)$ which is analogous to the effective potential vector given in the nonrelativistic regime in \cite{lin}, and in the relativistic regime in \cite{bf5,b4}. We also have that the external magnetic field given in (\ref{6}) produces no torque $\vec{\tau}=\vec{d}\times\vec{E}$ on the analogue of the electric dipole moment $\vec{d}=g\,\vec{b}$, satisfies the electrostatic conditions, and yields an uniform magnetic field along the $z$ direction given by $\vec{B}_{\mathrm{eff}}=\vec{\nabla}\times\left(\vec{b}\times\vec{B}\right)=\lambda_{m}\,b^{3}\,\hat{z}$. Hence, all condition for achieving the relativistic Landau-He-McKellar-Wilkens quantization are satisfied in the Lorentz symmetry violation background (\ref{6}). In the following, we solve the Dirac equation (\ref{5}) by using the definitions given in (\ref{6}), and write Eq. (\ref{5}) in the form:
\begin{eqnarray}
i\frac{\partial\psi}{\partial t}=m\hat{\beta}\psi-i\hat{\alpha}^{1}\left[\frac{\partial}{\partial\rho}+\frac{1}{2\rho}\right]\psi-i\frac{\hat{\alpha}^{2}}{\rho}\,\frac{\partial\psi}{\partial\varphi}-i\hat{\alpha}^{3}\frac{\partial\psi}{\partial z}-\frac{g\lambda_{m}b^{3}}{2}\,\rho\,\hat{\alpha}^{2}\,\psi.
\label{7}
\end{eqnarray} 

We can write the solution of the Dirac equation (\ref{7}) in the form: $\psi=e^{-i\mathcal{E}t}\left(\phi\,\,\,\chi\right)^{T}$, where $\phi$ and $\chi$ are two-spinors. Substituting this solution into (\ref{7}), we obtain two-coupled equation for $\phi$ and $\chi$. The first coupled equation is
\begin{eqnarray}
\phi=\frac{1}{\left(\mathcal{E}-m\right)}\left[-i\sigma^{1}\frac{\partial}{\partial\rho}-i\sigma^{1}\frac{1}{2\rho}-i\frac{\sigma^{2}}{\rho}\frac{\partial}{\partial\varphi}-i\sigma^{3}\frac{\partial}{\partial z}-\frac{g\lambda_{m}b^{3}}{2}\,\rho\,\sigma^{2}\right]\chi,
\label{8}
\end{eqnarray}
and the second coupled equation is
\begin{eqnarray}
\chi=\frac{1}{\left(\mathcal{E}+m\right)}\left[-i\sigma^{1}\frac{\partial}{\partial\rho}-i\sigma^{1}\frac{1}{2\rho}-i\frac{\sigma^{2}}{\rho}\frac{\partial}{\partial\varphi}-i\sigma^{3}\frac{\partial}{\partial z}-\frac{g\lambda_{m}b^{3}}{2}\,\rho\,\sigma^{2}\right]\phi.
\label{9}
\end{eqnarray}

Substituting $\chi$ in (\ref{9}) into (\ref{8}), we obtain the following second-order differential equation:
\begin{eqnarray}
\left(\mathcal{E}^{2}-m^{2}\right)\phi&=&-\frac{\partial^{2}\phi}{\partial\rho^{2}}-\frac{1}{\rho}\frac{\partial\phi}{\partial\rho}+\frac{1}{4\rho^{2}}\,\phi-\frac{1}{\rho^{2}}\,\frac{\partial^{2}\phi}{\partial\varphi^{2}}+\frac{i\sigma^{3}}{\rho^{2}}\,\frac{\partial\phi}{\partial\varphi}-\frac{\partial^{2}\phi}{\partial z^{2}}\nonumber\\
[-2mm]\label{10}\\[-2mm]
&+&ig\lambda_{m}b^{3}\frac{\partial\phi}{\partial\varphi}-\frac{g\lambda_{m}b^{3}}{2}\,\sigma^{3}\phi+\frac{\left(g\lambda_{m}b^{3}\right)^{2}}{4}\,\rho^{2}\phi.\nonumber
\end{eqnarray}

We can see in (\ref{10}) that $\phi$ is an eigenfunction of $\sigma^{3}$, whose eigenvalues are $s=\pm1$. Thus, we can write $\sigma^{3}\phi_{s}=\pm\phi_{s}=s\phi_{s}$. We can see that the operators $\hat{p}_{z}=-i\partial_{z}$ and $\hat{J}_{z}=-i\partial_{\varphi}$ \cite{schu} commute with the Hamiltonian of the right-hand side of (\ref{10}), therefore we can write the solution of (\ref{10}) in terms of the eigenvalues of the operator $\hat{p}_{z}=-i\partial_{z}$, and the $z$-component of the total angular momentum $\hat{J}_{z}=-i\partial_{\varphi}$ \footnote{The discussion about the expression of the $z$-component of the total angular momentum operator in cylindrical coordinates was done in Ref. \cite{schu}. There, it has been shown that the $z$-component of the total angular momentum in cylindrical coordinates is given by $\hat{J}_{z}=-i\partial_{\varphi}$, where the eigenvalues are $\mu=l\pm\frac{1}{2}$.}: 
\begin{eqnarray}
\phi_{s}=e^{i\left(l+\frac{1}{2}\right)\varphi}\,e^{ikz}\,G_{s}\left(\rho\right),
\label{11}
\end{eqnarray}
where $l=0,\pm1,\pm2,\ldots$ and $k$ is a constant. Substituting (\ref{11}) into the Schr\"odinger-Pauli equation (\ref{10}), we obtain
\begin{eqnarray}
G_{s}''+\frac{1}{\rho}G_{s}'-\frac{\nu_{s}^{2}}{\rho^{2}}G_{s}-\frac{\bar{\alpha}^{2}}{4}\,\rho^{2}\,G_{s}+\mu_{s}\,G_{s}=0,
\label{12}
\end{eqnarray}
where we have defined the parameters:
\begin{eqnarray}
\nu_{s}&=&l+\frac{1}{2}\left(1-s\right);\nonumber\\
\mu_{s}&=&\mathcal{E}^{2}-m^{2}-k^{2}+\bar{\alpha}\,\nu_{s}+s\bar{\alpha},\label{13}\\
\bar{\alpha}&=&g\lambda_{m}b^{3}.\nonumber
\end{eqnarray}

Our next step is to make a change of variables given in the form: $\eta=\frac{\bar{\alpha}}{2}\,\rho^{2}$. In this way, the radial equation (\ref{12}) becomes
\begin{eqnarray}
\eta\,G_{s}''+G_{s}'-\frac{\nu_{s}^{2}}{4\eta}G_{s}-\frac{\eta}{4}\,G_{s}+\frac{\mu_{s}}{2\bar{\alpha}}\,G_{s}=0.
\label{14}
\end{eqnarray}

In order to obtain a regular solution at the origin, we can write the solution of the second order differential equation (\ref{14}) in the form:
\begin{eqnarray}
G_{s}\left(\eta\right)=e^{-\frac{\eta}{2}}\,\eta^{\frac{\left|\nu_{s}\right|}{2}}\,M_{s}\left(\eta\right).
\label{15}
\end{eqnarray}

Substituting (\ref{15}) into (\ref{14}), we have
\begin{eqnarray}
\eta\,M_{s}''+\left[\left|\nu_{s}\right|+1-\eta\right]M_{s}''+\left[\frac{\mu_{s}}{2\bar{\alpha}}-\frac{\left|\nu_{s}\right|}{2}-\frac{1}{2}\right]M_{s}=0.
\label{16}
\end{eqnarray}
The second-order differential equation (\ref{16}) is called in the literature as the Kummer equation or the confluent hypergeometric equation \cite{abra}. The function $M_{s}\left(\eta\right)$ corresponds to the Kummer function of first kind which is defined as
\begin{eqnarray}
M_{s}\left(\eta\right)=M\left(\frac{\left|\nu_{s}\right|}{2}+\frac{1}{2}-\frac{\mu_{s}}{2\bar{\alpha}},\left|\nu_{s}\right|+1,\eta\right).
\label{17}
\end{eqnarray} 
In order to obtain a finite radial wave function, we must impose the condition where the confluent hypergeometric series becomes a polynomial of degree $n$ (where $n=0,1,2,\ldots$). This occurs when $\frac{\left|\nu_{s}\right|}{2}+\frac{1}{2}-\frac{\mu_{s}}{2\bar{\alpha}}=-n$. By taking the parameters defined in (\ref{13}), we have
\begin{eqnarray}
\mathcal{E}_{n,\,l,\,s}=\sqrt{m^{2}+k^{2}+2g\lambda_{m}b^{3}\left[n+\frac{\left|\nu_{s}\right|}{2}-\frac{\nu_{s}}{2}+\frac{1}{2}\left(1+s\right)\right]}.
\label{18}
\end{eqnarray}

The relativistic energy levels given in (\ref{18}) correspond to the relativistic Landau levels for the He-McKellar-Wilkens system \cite{lin,hmw} based on the Lorentz symmetry violation background. This analogue of the Landau quantization could be achieved by assuming that a Lorentz symmetry violation background defined by a fixed space-like vector field in the $z$ direction and the radial magnetic field produced by a magnetic charge density as suggested in Ref. \cite{lin}. 
%In this case, we have seen that the interaction between the electric dipole moment $\vec{d}=g\,\vec{b}$ and the radial magnetic field satisfies all con
Note that a change in the Lorentz symmetry violation scenario produces a change in the direction of the electric dipole moment or does not allow us to define the electric dipole moment for a Dirac neutral particle (for instance, if we have a fixed time-like vector $b^{\mu}=\left(b^{0},0,0,0\right)$) \cite{bbs}. In both cases, the relativistic Landau-He-McKellar-Wilkens quantization cannot be achieved via Lorentz symmetry breaking effects.

Moreover, we can also obtain the nonrelativistic limit of the energy levels (\ref{18}) by applying the Taylor expansion up to the first-order terms. Thereby, the Taylor expansion yields
\begin{eqnarray}
\mathcal{E}_{n,\,l}\approx m+\frac{g\lambda_{m}b^{3}}{m}\left[n+\frac{\left|\nu_{s}\right|}{2}-\frac{\nu_{s}}{2}+\frac{1}{2}\left(1+s\right)\right]+\frac{k^{2}}{2m},
\label{19}
\end{eqnarray}
where the first term of the right-hand-side of Eq. (\ref{19}) corresponds to the rest mass of the neutral particle. The remaining terms of  (\ref{19}) correspond to energy levels of the nonrelativistic Landau-He-McKellar-Wilkens quantization based on the Lorentz symmetry violation background.

\section{Coulomb-like potential induced by Lorentz symmetry breaking effects}

In this section, we solve the Dirac equation exactly in a background having a Coulomb-like potential induced by effects of the Lorentz symmetry breaking. Now, let us consider a fixed space-like vector field and a uniform magnetic field given by
\begin{eqnarray}
\vec{b}=b\,\hat{\rho};\,\,\,\,\,\vec{B}=B_{0}\,\hat{z},
\label{2.1}
\end{eqnarray}
where $\hat{\rho}$ and $\hat{z}$ correspond to unit vectors in the radial direction and $z$ direction, respectively. Note that the presence of this uniform magnetic field and the choice of a fixed space-like vector field being parallel to the radial direction gives rise to a new possible scenario for the measurement of the Lorentz symmetry breaking. In a recent work \cite{bbs3}, a Coulomb-like potential is yielded in the nonrelativistic limit of the Dirac equation by a term which plays the role of a scalar potential. This scalar potential arises from the interaction between a radial electric field and a fixed space-like vector field in the radial direction. In the present case, we show that a Coulomb-like potential is induced by Lorentz symmetry effects in the radial equation through the term $g\,\left(\vec{b}\times\vec{B}\right)$, which acts on the neutral particle as a vector potential. In the following, we show that the Dirac equation can be solved exactly in the Lorentz symmetry violation background defined in (\ref{2.1}), and relativistic bound states can be achieved in analogous way to having a Dirac equation under the influence of a Coulomb-like potential. Hence, the Dirac equation describing the interaction between a spin-half neutral particle and a uniform magnetic field in the Lorentz symmetry violation background becomes
\begin{eqnarray}
i\frac{\partial\psi}{\partial t}=m\hat{\beta}\,\psi-i\hat{\alpha}^{1}\left[\frac{\partial}{\partial\rho}+\frac{1}{2\rho}\right]\psi-i\frac{\hat{\alpha}^{2}}{\rho}\,\frac{\partial\psi}{\partial\varphi}-i\hat{\alpha}^{3}\frac{\partial\psi}{\partial z}+gbB_{0}\,\hat{\alpha}^{2}\,\psi.
\label{2.2}
\end{eqnarray}

Again, we can write the solution of the Dirac equation (\ref{2.2}) in the form: $\psi=e^{-i\mathcal{E}t}\left(\phi\,\,\,\chi\right)^{T}$, where $\phi$ and $\chi$ are two-spinors. Substituting this solution into (\ref{2.2}), we obtain two-coupled equation for $\phi$ and $\chi$. The first coupled equation is
\begin{eqnarray}
\left(\mathcal{E}-m\right)\phi=\left[-i\sigma^{1}\frac{\partial}{\partial\rho}-i\sigma^{1}\frac{1}{2\rho}-i\frac{\sigma^{2}}{\rho}\frac{\partial}{\partial\varphi}-i\sigma^{3}\frac{\partial}{\partial z}+gbB_{0}\,\sigma^{2}\right]\chi,
\label{2.3}
\end{eqnarray}
and the second coupled equation is
\begin{eqnarray}
\left(\mathcal{E}+m\right)\chi=\left[-i\sigma^{1}\frac{\partial}{\partial\rho}-i\sigma^{1}\frac{1}{2\rho}-i\frac{\sigma^{2}}{\rho}\frac{\partial}{\partial\varphi}-i\sigma^{3}\frac{\partial}{\partial z}+gbB_{0}\,\sigma^{2}\right]\phi.
\label{2.4}
\end{eqnarray}

By eliminating $\chi$ in (\ref{2.4}) and substituting it into (\ref{2.3}), we obtain the following second-order differential equation:
\begin{eqnarray}
\left(\mathcal{E}^{2}-m^{2}\right)\phi&=&-\frac{\partial^{2}\phi}{\partial\rho^{2}}-\frac{1}{\rho}\frac{\partial\phi}{\partial\rho}+\frac{1}{4\rho^{2}}\,\phi-\frac{1}{\rho^{2}}\,\frac{\partial^{2}\phi}{\partial\varphi^{2}}+\frac{i\sigma^{3}}{\rho^{2}}\,\frac{\partial\phi}{\partial\varphi}\nonumber\\
[-2mm]\label{2.5}\\[-2mm]
&-&i\frac{2gbB_{0}}{\rho}\frac{\partial\phi}{\partial\varphi}+\left(gbB_{0}\right)^{2}\phi-\frac{\partial^{2}\phi}{\partial z^{2}}.\nonumber
\end{eqnarray}

By following the steps from Eq. (\ref{10}) to Eq. (\ref{13}), we have that $\phi_{s}$ given in (\ref{11}) is also a solution for Eq. (\ref{2.5}), but the function $G_{s}\left(\rho\right)$ becomes the solution of the following radial equation:
\begin{eqnarray}
G_{s}''+\frac{1}{\rho}\,G_{s}'-\frac{\nu_{s}^{2}}{\rho^{2}}\,G_{s}-\frac{\delta}{\rho}\,G_{s}+\zeta^{2}\,G_{s}=0,
\label{2.6a}
\end{eqnarray}
where we have defined the parameters in (\ref{2.6a}):
\begin{eqnarray}
\nu_{s}&=&l+\frac{1}{2}\left(1-s\right);\nonumber\\
\delta&=&2gbB_{0}\nu_{s}+sgbB_{0};\label{2.7}\\
\zeta^{2}&=&\mathcal{E}^{2}-m^{2}-k^{2}-\left(gbB_{0}\right)^{2}.\nonumber
\end{eqnarray}

 Now, let us discuss the asymptotic behavior of the radial equation (\ref{2.6a}). For $\rho\rightarrow\infty$, we have 
\begin{eqnarray}
G_{s}''+\zeta^{2}\,G_{s}=0;
\label{2.8}
\end{eqnarray}
thus, we can find either scattering states $\left(G_{s}\cong e^{i\zeta\rho}\right)$, or bound states $\left(G_{s}\cong e^{-\tau\rho}\right)$ if we consider $\zeta^{2}=-\tau^{2}$ \cite{mello}. Note that the fourth term of the left-hand-side of Eq. (\ref{2.6a}) arises from the effects of the Lorentz symmetry breaking, and plays the role of a Coulomb-like potential. Our interest in this work is to obtain bound states, thus, the term proportional to $\delta$ behaves like an attractive potential by taking negative values of $\delta$, that is, by considering $\delta=-\left|\delta\right|$ \cite{mello}. This is possible for $g<0$ and $B_{0}>0$ (we consider $b$ being always a positive number) or $g>0$ and $B_{0}<0$. Hence, both these choices yield an analogue of an attractive Coulomb potential. Thereby, we consider $\zeta^{2}=-\tau^{2}$ and $\delta=-\left|\delta\right|$ from now on, and we rewrite Eq. (\ref{2.6a}) in the form:
\begin{eqnarray}
G_{s}''+\frac{1}{\rho}G_{s}'-\frac{\nu_{s}^{2}}{\rho^{2}}G_{s}+\frac{\left|\delta\right|}{\rho}\,G_{s}-\tau^{2}\,G_{s}=0.
\label{2.9}
\end{eqnarray}

Next, we make a change of variables given by $r=2\tau\rho$, then, the radial equation (\ref{2.9}) becomes
\begin{eqnarray}
G_{s}''+\frac{1}{r}G_{s}'-\frac{\nu_{s}^{2}}{r^{2}}\,G_{s}+\frac{\left|\delta\right|}{2\tau r}\,G_{s}-\frac{1}{4}\,G_{s}=0.
\label{2.10}
\end{eqnarray}
In order to obtain a regular solution at the origin for the second-order differential equation (\ref{2.10}), we can write
\begin{eqnarray}
G_{s}\left(r\right)=e^{-\frac{r}{2}}\,r^{\left|\nu_{s}\right|}\,M_{s}\left(r\right).
\label{2.11}
\end{eqnarray}
Substituting (\ref{2.11}) into (\ref{2.10}), we obtain the second-order differential equation:
\begin{eqnarray}
r\,M_{s}''+\left[2\left|\nu_{s}\right|+1-r\right]M_{s}'+\left[\frac{\left|\delta\right|}{2\tau}-\left|\nu_{s}\right|-\frac{1}{2}\right]M_{s}=0.
\label{2.12}
\end{eqnarray}

The second-order differential equation (\ref{2.12}) is known in the literature as the Kummer equation or the confluent hypergeometric equation \cite{abra}. The function $M_{s}\left(r\right)$ corresponds to the Kummer function of first kind which is defined as
\begin{eqnarray}
M_{s}\left(r\right)=M\left(\left|\nu_{s}\right|+\frac{1}{2}-\frac{\left|\delta\right|}{2\tau},2\left|\nu_{s}\right|+1,r\right).
\label{2.13}
\end{eqnarray} 
In order to obtain a finite radial wave function, we must impose the condition where the confluent hypergeometric series becomes a polynomial of degree $n$ (where $n=0,1,2,\ldots$). This occurs when $\left|\nu_{s}\right|+\frac{1}{2}-\frac{\left|\delta\right|}{2\tau}=-n$. In this way, with $\left(-\tau^{2}\right)=\mathcal{E}^{2}-m^{2}-k^{2}-\left(gbB_{0}\right)^{2}$, we obtain
\begin{eqnarray}
\mathcal{E}_{n,\,l}=\sqrt{m^{2}+k^{2}+\left(gbB_{0}\right)^{2}-\frac{\left[2gbB_{0}\nu_{s}+sgbB_{0}\right]^{2}}{4\left[n+\left|\nu_{s}\right|+1/2\right]^{2}}}.
\label{2.14}
\end{eqnarray}

Hence, we have that the energy levels (\ref{2.14}) correspond to the relativistic energy levels for a neutral particle under the influence of a Coulomb-like potential induced by Lorentz symmetry breaking effects. We have seen that this Coulomb-like potential is produced by a different Lorentz symmetry violation scenario in relation to that of the previous section. In this case, the Lorentz symmetry violation scenario is defined by a fixed space-like vector parallel to the radial direction. Thereby, by considering the presence of a uniform magnetic field along the $z$-axis, we have seen that the Dirac equation can be solved exactly, and relativistic bound states analogous to having a Dirac neutral particle under the influence of an attractive Coulomb potential can be obtained.

Finally, we can obtain the nonrelativistic limit of the energy levels (\ref{2.14}) by applying the Taylor expansion up to the first-order terms. Then, the Taylor expansion yields
\begin{eqnarray}
\mathcal{E}_{n,\,l}\approx m-\frac{1}{8m}\frac{\left[2gbB_{0}\nu_{s}+sgbB_{0}\right]^{2}}{\left[n+\left|\nu_{s}\right|+1/2\right]^{2}}+\frac{\left(gbB_{0}\right)^{2}}{2m}+\frac{k^{2}}{2m},
\label{2.15}
\end{eqnarray}
where the first term of the right-hand-side of Eq. (\ref{2.15}) corresponds to the rest mass of the neutral particle. The remaining terms of  (\ref{2.15}) correspond to the nonrelativistic energy levels for a neutral particle under the influence of a Coulomb-like potential induced by the effects of the Lorentz symmetry violation background.

\section{conclusions}

In this work, we have discussed the relativistic Landau-He-McKellar-Wilkens quantization induced by Lorentz symmetry breaking effects. This analogue of the Landau quantization is achieved by considering an electric dipole moment defined in terms of a fixed space-like vector field in the $z$ direction and the radial magnetic field suggested in Ref. \cite{lin}, produced by a magnetic charge density. From this scenario, we have seen that the electrostatic conditions, the absence of torque on the electric dipole moment, and the presence of a uniform effective magnetic field are satisfied, therefore the relativistic Landau-He-McKellar-Wilkens quantization based on this Lorentz symmetry violation scenario can be obtained.

We have also seen, by changing the Lorentz symmetry violation scenario, that we can discuss the behaviour of a Dirac neutral particle under the influence of a Coulomb-like potential induced by Lorentz symmetry breaking effects. In this case, we have considered a Lorentz symmetry violation scenario given by a fixed space-like vector parallel to the radial direction and the presence of a uniform magnetic field along the $z$ direction. Hence, we have shown that the Dirac equation can be solved exactly in this new Lorentz symmetry violation scenario, and relativistic bound states for a Dirac neutral particle under the influence of a Coulomb-like potential (induced by Lorentz symmetry breaking effects) can be achieved.

The problem of detecting Lorentz violating effects at low energies (until the electroweak unification scale) is that the modulus of the 4-vector background is very tenuous. Therefore, what has been made in recent years is to establish limits of energy in which this breaking can or cannot be seen \cite{belich,belich1,belich2,belich3}.

An interesting point of discussion that deserves to be commented is the recent studies of Dirac-like systems, which have attracted a great deal of attention \cite{l12,bf30,l13}. As example, the Landau quantization has been discussed in graphene \cite{l13} in the presence of topological defects \cite{l12}, and a model for confining a Dirac particle in a two-dimensional quantum ring and a quantum dot has been proposed in \cite{bf30}. Based on the discussion about the influence of Lorentz symmetry breaking effects on the hydrogen atom \cite{Nonmini} and geometric quantum phases \cite{belich1,belich,bbs2,bbs3,bb}, Dirac-like systems \cite{l13,l12,bf30} can open new possible scenarios of studying Lorentz symmetry breaking effects.

We would like to thank CNPq (Conselho Nacional de Desenvolvimento Cient\'ifico e Tecnol\'ogico - Brazil) for financial support.

\end{document}